\documentclass{article}
\usepackage{graphicx,amsmath}
\usepackage{a4wide}

\newcommand{\ee}{\end{enumerate}}        
\newcommand{\bi}{\begin{itemize}}
\newcommand{\ei}{\end{itemize}}        
\newcommand{\eq}[1]{{\protect\frenchspacing eq.~(\ref{#1})}}
\newcommand{\fig}[1]{{\protect\frenchspacing Fig.~\ref{#1}}}
\newcommand{\reff}[1]{{\frenchspacing Ref.~\cite{#1}}}
\newcommand{\tab}[1]{Tab.~\ref{#1}}
\newcommand{\beq}{\protect\begin{equation}}
\newcommand{\eeq}{\protect\end{equation}}        
\newcommand{\bqa}{\begin{eqnarray}}        
\newcommand{\eqa}{\end{eqnarray}}        
\newcommand{\be}{\begin{enumerate}}
\newcommand{\tauflip}{\tau_{\mbox{\tiny flip}}}
\newcommand{\taueff}{\tau_{\mbox{\tiny eff}}}
\newcommand{\pmuca}{P_{\mbox{\tiny MUCA}}}
\newcommand{\vmuca}{V_{\mbox{\tiny MUCA}}}
\newcommand{\wmuca}{W_{\mbox{\tiny MUCA}}}
\newcommand{\nmuca}{N_{\mbox{\tiny MUCA}}}
\newcommand{\Nts}{N_{\mbox{\tiny ts}}}
\newcommand{\Nindep}{N_{\mbox{\tiny indep}}}
\newcommand{\sigmamuca}{\sigma_{\mbox{\tiny MUCA}}}
\newcommand{\sigmacan}{\sigma_{\mbox{\tiny can}}}

\newcommand{\ie}{{\frenchspacing i.\hspace{0.4mm}e.{}}}
\newcommand{\nn}{\nonumber}

\newcommand{\prl}[3]{\frenchspacing Phys. Rev. Lett. {\bf #1} {(#2)}
  {#3}}
\newcommand{\plb}[3]{ \frenchspacing Phys.  Lett.  {\bf #1B} {(#2)}
  {#3}}

\newcommand{\prd}[3]{ \frenchspacing Phys.  Rev.   {D\bf #1} {(#2)}
  {#3}}
\newcommand{\pre}[3]{ \frenchspacing Phys.  Rev.   {Eg\bf #1} {(#2)}
  {#3}}
\newcommand{\jcp}[3]{ \frenchspacing J.  Comp. Phys.   {\bf #1} {(#2)}
  {#3}}
\newcommand{\jsp}[3]{ \frenchspacing J.  Stat. Phys.   {\bf #1} {(#2)}
  {#3}}
\newcommand{\jcep}[3]{ \frenchspacing J.  Chem. Phys.   {\bf #1} {(#2)}
  {#3}}

\newcommand{\npb}[3]{ \frenchspacing Nucl.  Phys. {\bf B#1} {(#2)}
  {#3}}
\newcommand{\npbfs}[4]{ \frenchspacing Nucl. Phys. {\bf B#1} {FS[#2]}
  {(#3)} {#4}}
\newcommand{\npps}[3]{\frenchspacing Nucl.  Phys. B (Proc. Suppl.)
  {\bf #1} {(#2)} {#3}}

\title{
{\small
\hspace*{\fill} HLRZ-1998-56\\
\hspace*{\fill} WUB-98/32\\}
  Multicanonical Hybrid Monte Carlo:\\ Boosting Simulations of Compact
  QED}

\author{
  G.~Arnold, K.~Schilling\\
  {\small NIC, c/o Research Center J\"{u}lich and DESY, Hamburg,
    D-52425
    J\"{u}lich, Germany}\\[8pt]
  {Th.~Lippert}\\
 {\small Department of Physics, University of
    Wuppertal, D-42097 Wuppertal, Germany}}

\date{ }

\begin{document}

\maketitle

\begin{abstract}
  We demonstrate that substantial progress can be achieved in the
  study of the phase structure of 4-dimensional compact QED by a joint
  use of hybrid Monte Carlo and multicanonical algorithms, through an
  efficient parallel implementation. This is borne out by the
  observation of considerable speedup of tunnelling between the
  metastable states, close to the phase transition, on the Wilson
  line. We estimate that the creation of adequate samples (with order
  100 flip-flops) becomes a matter of half a year's runtime at 2
  Gflops sustained performance for lattices of size up to $24^4$.
\end{abstract}


\section{Introduction}

It appears exceedingly intriguing to define variants of QED by
studying vacuum states other than the usual perturbative vacuum.
Lattice techniques have the potential power to deal with this
situation, whenever they provide us with phase transition points of
second and higher orders.

It is embarrassing that lattice simulations of compact QED still have
not succeeded to clarify the order of the phase transition near $\beta
= 1$, the existence of which was established in the classical paper of
Guth \cite{GUT78}. This is mainly due to the failure of standard
updating algorithms, like metropolis, heatbath or metropolis with
reflections \cite{MET53,REB79,ADL81,BUN94}, to move the system at
sufficient rate between the observed metastable states near its phase
transition.  The tunneling rates decrease exponentially in $L^3$ and
exclude the use of lattices large enough to make contact with the
thermodynamic limit by finite size scaling techniques (FSS)
\cite{FIS70}.

In this paper, we propose to make use of the multicanonical (MUCA)
algorithm \cite{BER91} within the hybrid Monte Carlo (HMC) updating
scheme \cite{DUA87} in order to boost the tunneling rates.  Since both
algorithms are inherently of global nature, their combination will
facilitate the parallelization of MUCA which could not be achieved
otherwise.

In the early days of simulations on the hyper-torus, the U(1) phase
transition was claimed to be second-order
\cite{LAU80,BHA81,MUE82,GUP86}, however, with increasing lattice
sizes, metastabilities and double peak action distributions became
manifest, strongly hinting at its first-order character
\cite{JER83,AZC91,BHA92}. This picture is in accord with various
renormalization group studies \cite{LAN87,HAS88}.

However, the latent heat is found to decrease with the lattice size
and the critical exponent $\nu$ is neither 0.25 (first order) nor
$0.5$ (trivially second-order) \cite{KLA97}. These facts allow for two
possible propositions:
\begin{enumerate}
\item the double peak structure is a finite size effect and might
  vanish in the thermodynamic limit, leading to the signature of a
  second-order phase transition;
\item the phase transition is weakly first-order, \ie, the correlation
  length $\xi$ is finite, but large in terms of the available lattice
  extent $L$; this would fake, on small lattices, the signature of a
  second-order transition, and a stabilized value of the latent heat
  would only become visible in the thermodynamic limit $L>\xi$ .
\end{enumerate}

In search for a lattice formulation of QED with a second-order
transition point, the action was generalized to include a piece in the
adjoint representation with coupling $\gamma$ \cite{BHA81-2}, in the
expectation that the phase transition would be driven towards
second-order, at sufficiently small negative values of $\gamma$.
However, simulations on the hyper-torus, up to $\gamma=-0.4$, revealed
the reappearance of a double peak on large enough lattices
\cite{CAM97,CAM98}.  Thus, the hypothesis of a second-order phase
transition at some finite negative value of $\gamma$
\cite{EVE85,JER97,JER97-2} is again doubted; furthermore,
renormalization group investigations indicated that the second-order
phase transition is located at $\gamma\rightarrow -\infty$
\cite{HAS88}.

Ref. \cite{JER85} has speculated that the mechanism behind the lattice
heuristics of metastabilities is driven by monopole loops that wrap
around the hyper-torus. According to this scenario, the inefficiency
of local updating algorithms to create and annihilate such monopole
constellations causes their slowing down, in agreement with the
earlier proposition (1).  Results were presented in support of this
view by switching to spherical lattices with trivial homotopy group
where such wrapping loops are no more topologically stabilized
\cite{LAN94,LAN94-2,JER95, BAI94}.  But on spherical lattices
equivalent to $L=26$ at $\gamma=-0.2$, double peak structures have
recently been reported to reappear \cite{CAM97,CAM98}, corroborating
earlier observations with periodic boundary conditions at $\gamma=0$:
the suppression of monopole loop penetration through the lattice
surface turned out to be incapable of preventing the incriminated
double peak signal to show up on large lattices, to say $L=32$
\cite{LIP94,LIP95,LIP95-2}.

It appears that a clarification of the situation on the Wilson line is
mandatory for further progress in the understanding of compact lattice
QED! This challenge requires the design of more powerful updating
algorithms.  A promising method is based on simulated tempering
\cite{MAR92,KER94,KER94-2}, enlarging the Lagrangian by a monopole
term whose coupling is treated as an additional dynamical variable.
Multi-scale update schemes in principle can alleviate critical slowing
down (CSD) which is associated with the increase of the correlation
length $\xi$ (as measured in a non-mixed phase) near the critical
coupling, $\beta_c$ \cite{ADL91}.  However, the exponential
supercritical slowing down (SCSD) which is a consequence of the
surface tension at first-order phase transitions cannot be overcome by
such type of scale-adapted methods. In that instance, one expects the
autocorrelation times to grow exponentially with the system size due
to the occurrence of two 3-dimensional interfaces, leading to
\begin{equation}
  \tau_{\mbox{\tiny SCSD}}\propto
  \exp\left(2\sigma L^3 \right).
\label{EQ:SCSD}
\end{equation}

Torrie and Valleau \cite{TORRIE76,TORRIE762,VALLEAU90,DING92} have
shown how to generate arbitrary ``non-physical'' sampling
distributions. Their method, termed ``umbrella sampling'', has been
introduced to span large regions of phase diagrams. The method is
capable to improve the efficiency of stochastic sampling for
situations when dynamically nearly disconnected parts of phase space
occur by biassing the system to frequent the dynamically depleted,
connecting regions of configuration space. They interpreted their
method as sampling ``a whole range of temperatures'' \cite{TORRIE76}.

In recent years the idea of ``umbrella sampling'' has been popularized
and extensively applied under the name ``multi-canonical algorithm''
(MUCA) by Berg and Neuhaus \cite{BER91,BER91-2,BER95,BER97,BER97-2} to
the simulation of a variety of systems exhibiting first-order phase
transitions \cite{GRO92,RUM92,JAN93,JAN93-2,HAN94,NEU96}.  In this
procedure, the biassing weight $w(S)$ of a configuration with action
$S$ is dynamically adjusted (bootstrapped) such as to achieve a
near-constant overall frequency distribution over a wide range of $S$
within a {\em single} simulation.

Obviously, MUCA in principle offers a powerful handle to deal with
SCSD.  It remains then a practical question whether one can indeed
proceed to large lattices by boosting tunneling rates from the SCSD
behaviour (\eq{EQ:SCSD}) to the peak efficiency of local Monte Carlo
methods (characterized by ${O}((L^4)^2)$ complexity). This leads us to
the key point of this paper: it is a severe shortcoming of the
multi-canonical algorithm that its implementation is seemingly
restricted to sequential computers, as it requires knowledge of the
{\it global} action, even during local updating.  We will show that
HMC is from the very outset able to implement MUCA in a parallel
manner.

In sections \ref{SEC:MC} and \ref{SEC:HMC}, we will give a short
review of the MUCA and the HMC algorithms.  In section
\ref{SEC:MERGE}, we merge MUCA with HMC.  From our ongoing simulation
project of U(1) theory on the Wilson line \cite{ARN98}, we determine
the tunneling efficiency compared to the standard metropolis algorithm
which in our case is complemented by three reflection steps.  In
section \ref{SEC:RES}, we shall present our results for lattice sizes
up to $16^4$ and predict the tunnelling rates for lattice sizes up to
$24^4$, as would be required for a proper FSS.

\section{Multicanonical Hybrid Monte Carlo}

The hybrid Monte Carlo (HMC) algorithm \cite{DUA87,GOT87,LIP97}
produces a global trial configuration by carrying out a molecular
dynamics evolution of the field configuration very close to the
surface of constant action. Subsequently, a Monte Carlo decision is
imposed which is based on the global action difference $\Delta S$,
being small enough to be frequently accepted.  Within HMC, all
degrees-of-freedom can be changed simultaneously and hence in
parallel.  This then provides the straightforward path to implement
MUCA as part of HMC on parallel machines\footnote{A first attempt in
  this direction has been made in \reff{SEYFRIED} in the framework of
  the Higgs-Yukawa model.}: one just uses the values of the global
action, as provided by HMC, to compute the bias function $w(S)$ for
the MUCA algorithm.

\subsection{Multicanonical Algorithm\label{SEC:MC}}

``Canonical'' Monte Carlo generates a sample of field configurations,
$\left\{ \phi\right\}$, within a Markov process, according to the
Boltzmann weight,
\begin{equation}
  P_{\beta}(S)=\frac{1}{Z_{\mbox{\tiny can}}} e^{-\beta
    S(\phi)},
\end{equation}
which follows from maximizing the entropy with respect to all possible
probability distributions $P[\phi]$.  The partition function $Z$
normalizes the total probability to 1,
\begin{equation}
  Z_{\beta}=\int [d\phi]\,e^{-\beta S[\phi]}.
\end{equation}
$S$ is the action (the energy in the case of statistical mechanics)
and $\beta$ the coupling (or inverse temperature $\frac{1}{kT}$).

The canonical action density which in general exhibits a double peak
structure  at a first-order phase transition, can be rewritten as
\begin{equation}
  N_{\mbox{\tiny can}}(S,\beta)=\rho(S)\, e^{-\beta S},
\label{CAD}
\end{equation}
with the spectral density $\rho(S)$ being independent of $\beta$.
Usually, $ \int N_{\mbox{\tiny can}}(S,\beta)\,dS$ is set to 1.

The multicanonical approach aims at generating a flat action density
\begin{equation}
  N_{\mbox{\tiny MUCA}}(S,\beta)=\mbox{const., \hspace{.5cm} for }
  S_{\mbox{\tiny min}}\le S\le S_{\mbox{\tiny max}},
\end{equation}
in a range of $S$ that covers the double peaks at the first-order
phase transition.

Therefore, instead of sampling canonically according to $e^{-\beta
  S}$, one modifies  the sampling by a weight factor $\wmuca$:
\begin{equation}
\wmuca(S,\beta)=\left\{
\begin{array}{rll}
  \frac{1}{\rho( S_{\mbox{\tiny min}})} & e^{\beta S_{\mbox{\tiny
        min}}} &\qquad\mbox{for}\; S<S_{\mbox{\tiny min}}\\
  \frac{1}{\rho(S)} & e^{\beta S} &\qquad\mbox{for}\; S_{\mbox{\tiny
      min}}\le
  S\le S_{\mbox{\tiny max}}\\
  \frac{1}{\rho( S_{\mbox{\tiny max}})} & e^{\beta S_{\mbox{\tiny
        max}}} &\qquad \mbox{for}\; S>S_{\mbox{\tiny max}},\\
\end{array}\right. 
\label{WF}
\end{equation}
which is constant outside the relevant action range. Such `corrigez la
fortune' is equivalent to a net sampling according to the {\em yet}
unknown probability distribution
\begin{equation}
  w(S)=\frac{1}{\rho(S)},\qquad \mbox{for}\; S_{\mbox{\tiny min}}\le
  S\le S_{\mbox{\tiny max}}.
\end{equation}
Since $W(S,\beta)$ is unknown at the begin of the simulation, it is
instrumental for MUCA to follow M\"unchhausen and bootstrap from good
guesstimates \cite{BER97}.  We shall do so by starting from an
observed histogram of the canonical action density,
$\hat{N}_{\mbox{\tiny can}}(S,\hat\beta_c)$, see \eq{CAD}, at the
supposed location of the phase transition\footnote{Hatted quantities
  refer to stochastic estimates.}, $\hat\beta_c$.  From the action
density, we compute $\hat\wmuca(S,\hat\beta_c)$ according to \eq{WF}.
The sampling then proceeds with the full MUCA weight,
\begin{eqnarray}
  \hat\pmuca(S)&\propto& e^{-\hat \beta_c S}
  \hat\wmuca(S,\hat\beta_c)\nn\\
&\propto& e^{-\big(\hat\beta_c+\hat\beta(S)\big) S -\hat\alpha(S)}.
\label{Pmuca}
\end{eqnarray}
This latter formulation can be interpreted as a simulation proceeding
at varying couplings (temperatures), hence the naming `{\em
  multicanonical}'.

In order to compute expectation values of observables $\cal O$, one
has to reweight the resulting action density at the end of the day by
the factor $\hat\wmuca(S,\hat\beta_c)$, which reconstitutes the proper
canonical density:
\begin{equation}
  \langle \hat{\cal O}_{\hat\beta_c}\rangle=\frac{ \sum_i {\cal
      O}^i_{\hat\beta_c}\,\frac{1}{\hat\wmuca(S_i,\hat\beta_c)}} {\sum_i\frac{1}{\hat
    \wmuca(S_i,\hat\beta_c)}}.
\label{AVE}
\end{equation}
Additionally, $\langle \hat{\cal O}_{\hat\beta_c}\rangle$ simulated at
$\hat\beta_c$ can be reweighted to any desired $\beta$ (following
\cite{SF}), given that the corresponding region of phase space has
been covered by the MUCA simulation. We emphasize that \eq{AVE} is
only useful complemented by a proper error analysis.  The canonical
error computed from the multicanonical ensemble has been elaborated in
Ref.~\cite{JAN94}.

Note that there are many possible choices for the form of the multicanonical
weight. Just for technical reasons we require it to be continuous in
$S$.  One can either guess an analytic function, or choose a polygonal
approximation such as given in \eq{Pmuca}.  In this case, the
multicanonical weight is expressed in terms of the functions
$\hat\beta(S)$ and $\hat\alpha(S)$ which are actually characteristic
functions of the bins.  $\hat\beta(S)$ can be considered as an
effective temperature \cite{TORRIE76}.

The computation of the weights requires the knowledge of the global
and not just the local change in action for each single update step.
For this reason, even for a local action, one cannot perform local
updating moves in parallel, such as the well-known checkerboard
pattern. As a consequence, MUCA is not parallelizable for local update
algorithms.  For remedy, we propose here to {\it go global} and
utilize the HMC updating procedure.

\subsection{Hybrid Monte Carlo\label{SEC:HMC}}

The HMC consists of two parts: the hybrid molecular dynamics algorithm
(HMD) evolves the degrees of freedom by means of molecular dynamics
(MD) which is followed by a global Metropolis decision to render the
algorithm exact.

In addition to the gauge fields $\phi_{\mu}(x)$ one introduces a set
of statistically independent canonical momenta $\pi_{\mu}(x)$, chosen
at random according to a Gaussian distribution
$\exp(-\frac{\pi^2}{2})$.  The action $S[\phi]$ is extended to a
guidance Hamiltonian \beq
\label{Hguidance}
{\cal H}[\phi,\pi]=\frac{1}{2}\sum_{\mu,x}\pi_\mu^2(x)+\beta S[\phi].
\end{equation}
Starting with a configuration $(\phi,\pi)$ at MD time $t=0$, the
system moves through phase space according to the equations of
motion
\bqa%
\label{EOM} \dot{\phi}_\mu &=& \frac{\partial {\cal H}}{\partial
  \pi_\mu} =
\pi_\mu, \nn \\
\dot{\pi}_\mu &=& -\frac{\partial {\cal H}}{\partial \phi_\mu}
=-\frac{\partial} {\partial \phi_\mu}[\beta S],
\eqa %
leading to a proposal configuration $(\phi^\prime,\pi^\prime)$ at time
$t=\tau$. Finally this proposal is accepted in a global Metropolis
step with probability
\begin{equation}
\label{Hacceptance}
P_{acc}=\min\left( 1,e^{-\Delta {\cal H}}\right), \quad \mbox{with} \ 
\Delta {\cal H}={\cal H}[\phi^\prime,\pi^\prime]-{\cal H}[\phi,\pi].
\end{equation}

The equations of motion are integrated numerically with finite step
size $\Delta t$ along the trajectory from $t=0$ up to
$t=N_{\mbox{\tiny md}}$.  Using the leap-frog scheme as symplectic
integrator the discretized version of \eq{EOM} reads:
\bqa%
\label{EOMdiscret}
\phi^{n+1} &=& \phi^n+\Delta t \cdot \pi^n-\frac{\Delta t^2}{2}
\frac{\partial}{\partial \phi}\Big(\beta S[\phi^n]\Big) \nn \\
\pi^{n+1} &=& \pi^n-\frac{\Delta t}{2} \frac{\partial}{\partial
  \phi}\Big(\beta S[\phi^n]\Big)-\frac{\Delta t}{2} \frac{\partial
  }{\partial \phi}\Big(\beta S[\phi^{n+1}]\Big).
\eqa%
Here we have presented the scheme with both the momenta and the gauge
fields defined at full time steps\footnote{Note that the actual
  implementation computes the momenta at half time steps according to
  the sequence
\bqa%
\pi(t+\Delta t/2) &=& \pi(t-\Delta t/2)-\Delta t\frac{\partial
  }{\partial \phi}\big(\beta S[\phi(t)]\big), \nn \\
\phi(t+\Delta t) &=& \phi(t)+\Delta t \cdot \pi(t+\Delta t/2), \nn
\eqa%
initialized and finished by a half-step in $\pi$ \cite{DUA87}. Each
sequence approximates the correct ${\cal H}$ with an error of
$O(\Delta t^3)$.}, $t=n \Delta t$.

To ensure that the Markov chain of gauge field configurations reaches
a {\em unique} fixed point distribution $\exp(-S[\phi])$ one must
require the updating procedure to fulfil {\em detailed balance},
which is guaranteed by the iterative map of \eq{EOMdiscret} $f:
(\phi,\pi)\to(\phi^\prime,\pi^\prime)$ being
\begin{itemize}
\item time reversible: $f(\phi^\prime,-\pi^\prime)=(\phi,-\pi)$
\item and measure preserving: $[d\phi^\prime][d\pi^\prime]=[d\phi][d\pi]$.
\end{itemize}

It is easy to prove that these two conditions also hold for the
multicanonical action. Note that the guidance Hamiltonian,
\eq{Hguidance}, defining the MD may differ from the acceptance
Hamiltonian in \eq{Hacceptance}, which produces the equilibrium
distribution proper.  In the following, we shall exploit this freedom
to develop two variant mergers of MUCA and HMC.

\subsection{Merging MUCA and HMC for Compact QED (MHMC)\label{SEC:MERGE}}

We consider a multicanonical HMC for pure 4-dimensional $U(1)$ gauge theory
with standard Wilson action defined as
\beq%
S[\phi]=\sum_{x,\nu>\mu}\Big[1-\cos\big(\theta_{\mu\nu}(x)\big)\Big],
\end{equation}
where
\[
\theta_{\mu\nu}(x)=\phi_\mu(x)+\phi_\nu(x+\hat{\mu})-\phi_\mu(x+\hat{\nu})-\phi_\nu(x)
\]
is the sum of link angles that contribute to one of six plaquettes
interacting with the link angle $\phi_\mu(x)$.

Eq.~\ref{Pmuca} suggests to consider an effective action $\hat{S}$
including the ``multicanonical potential'' $\vmuca$
\bqa%
\hat{S}&=&\hat{\beta}_c S+\hat\vmuca(S,\hat\beta_c),
\label{Seff}
\eqa%
with
\bqa%
\hat\vmuca(S,\hat\beta_c)&=&\log\Big({\frac{1}{\hat\wmuca(S,{\hat\beta}_c)}}\Big)\nn \\
&=& \hat{\beta}(S)\,S+\hat{\alpha}(S).
\eqa%
There are two natural options to proceed from here:
\begin{description}
\item[Method 1] performs molecular dynamics using the canonical
  guidance Hamiltonian 
\[
{\cal H}_{md}=\frac{1}{2}\sum\pi^2+{\hat \beta_c} S,
\] 
with standard action $S$.  The resulting gluonic force is given by
\bqa%
F(x)&=&\dot{\pi}_\mu(x)\nn\\
&=&{\hat \beta}_c \sum_{\nu\not= \mu}\Big[\sin\theta_{\mu\nu}(x-{\hat
  \nu})-\sin\theta_{\mu\nu}(x)\Big].
\eqa%
\item[Method 2] makes use of the multicanonical potential as a driving
  term within the Hamiltonian, 
\[
{\cal
      H}_{md}=\frac{1}{2}\sum\pi^2+\hat{S},
\] 
inducing an additional drift term
\bqa%
\dot{\pi}_\mu(x) &=&
F(x)-\frac{\partial}{\partial\phi_\mu(x)}\hat\vmuca\Big(S\big[\phi(x)\big],\hat\beta_c\Big)\nn \\
        &=& \Big({\hat\beta_c}+{\hat\beta(S)}\Big) \sum_{\nu\not=
\mu}\Big[\sin\theta_{\mu\nu}(x-{\hat \nu})-\sin\theta_{\mu\nu}(x)\Big],
\eqa%
with the effective $\hat\beta$ as defined in \eq{Pmuca}.
\end{description}
For both options, the Hamiltonian governing the accept/reject decision,
\eq{Hacceptance}, reads equally:
\begin{equation}
{\cal H}_{\mbox{\tiny acc}}=\frac{1}{2}\sum\pi^2+\hat{S}.
\end{equation}
The latter method is governed by the dynamics underlying the very two peak
structure: as one can see from \fig{plot16}, $\vmuca$ is repelling
the system out of the hot (cold) phase towards the cold (hot)
phase, thus increasing its mobility and enhancing flip-flop activity.

We comment that the implementation of method 2 requires the
computation of the global action (to adjust the correct multicanonical
weight, \eq{Pmuca}) at each integration step along the trajectory of
molecular dynamics to guarantee reversibility. In the polygon
approximation, this amounts to a determination of the effective
coupling $\hat{\beta}$ at each time step in MD. Note that
$\hat{\alpha}$ does not influence MD but enters into the global
Metropolis decision, \eq{Hacceptance}.  Method 1 is much simpler,
running at {\it fixed} trial coupling $\hat\beta_c$ and avoiding the
effort of computing $N_{\mbox{\tiny md}}-2$ global sums while
travelling along the trajectories.  It turned out that of both
versions of MHMC, method 2 performs better than method 1, when
autocorrelation and difference in computational effort are taken into
account. Thus, we continue our analysis by investigation of method 2.

\section{Results\label{SEC:RES}}

In order to evaluate the efficiency of the MHMC and to set the stage
for a proper extrapolation, we generated time series of the U(1)
action on lattices of size $6^4$ up to $16^4$.  The runs are
summarized in \tab{SERIES}. We applied method 2 as being the more
promising one on real large lattices.
\begin{table}[tb]
\begin{center}
\begin{tabular}{|r|l|r||r|r|r|}
\hline
L & $\beta$     & \# sweeps(MRS) & \# sweeps(MHMC) & $\Delta t$ & $N_{md}$\\ \hline \hline
6 & 1.001600    & 1.460.000  & 1.650.000 & 0.120 &10 \\ \hline
8 & 1.007370    & 1.320.000  & 1.430.000 & 0.093 &13 \\ \hline
10& 1.009300    & 1.030.000  &   560.000 & 0.071 &17 \\ \hline
12&  1.010150   &   680.000  &           & 0.060 &20    \\ \hline
  & 1.010143    & 1.790.000  & 1.160.000 & &  \\ \hline
14&1.010300     & 1.430.000  &           & 0.050 &24\\ \hline
  &1.010668     &   900.000  &   990.000 & &  \\ \hline
16&1.010800     & 1.210.000  &           & 0.045 & 26\\ \hline
  &1.010753     &   750.000  &   760.000 & & \\ \hline
\end{tabular}
\caption{Total numbers of
  sweeps carried out both for MRS and MHMC at different lattice sizes
  $L$ and couplings $\beta$.\label{SERIES}}
\end{center}
\end{table}

\begin{figure}[!htb]
\centerline{\includegraphics[width=.7\textwidth]{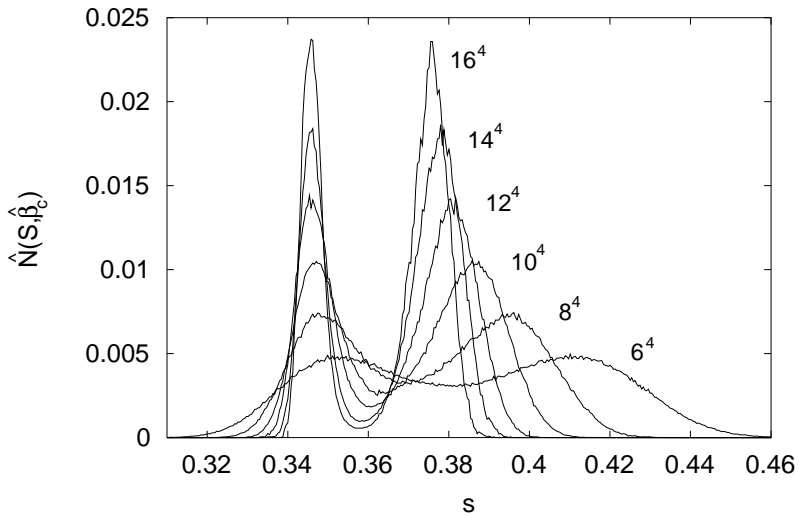}}
\caption{
  The canonical action densities, \eq{CAD}, on $6^4$ up to $16^4$
  lattices, reweighted to their respective $\hat \beta_c$ values, here
  defined via equal height of the histograms.
\label{WEIGHT2}}
\end{figure}

\begin{figure}[!htb]
\centerline{\includegraphics[width=.7\textwidth]{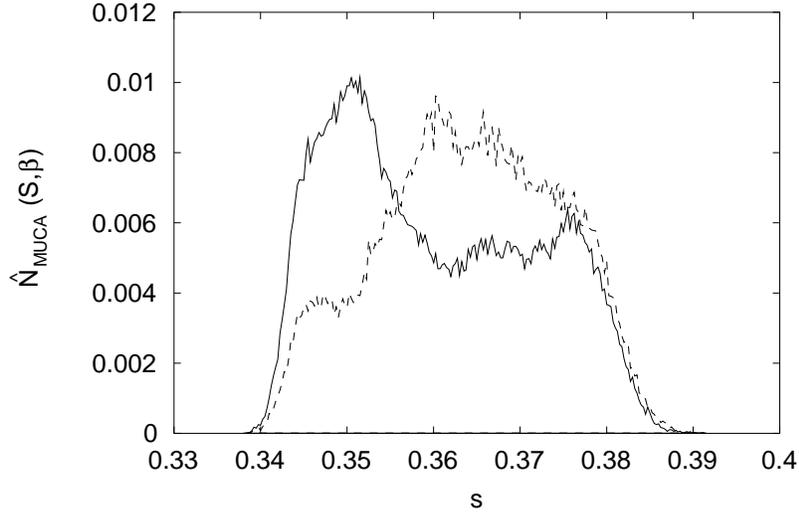}}
\caption{
  Evolution of multicanonical action density. Starting from a
  canonical run a first MHMC simulation is performed (dashed line),
  and, based on this result, the final run (solid line) is carried
  out.  Both curves are computed on the $16^4$ lattice at
  $\beta=1.010753$.
\label{WEIGHT}}
\end{figure}

To arrive at an action density $N_{\mbox{\tiny MUCA}}(S,\beta)$ which
is approximately flat in the desired region between the two peaks it
is crucial to find a good estimate $\hat{N}_{\mbox{\tiny
    can}}(S,\hat\beta_c)$ of the canonical action density.  However,
it becomes more and more delicate for large volumes to find the proper
multicanonical weight, $\hat\pmuca^L(S)$ (\eq{Pmuca}).  \fig{WEIGHT2}
shows the evolution of the action density as the lattice size
increases.
For lattices $<16^4$ it was sufficient to perform a short canonical
run to generate an action density $\hat\nmuca$ suitable to compute a
proper weight factor $\hat\wmuca$. On the $16^4$ system, however, the
resulting multicanonical distribution becomes quite sensitive to the
choice of the weight.  Therefore, in the case of large lattices ($L\ge
16$) we cannot rely on canonical simulations to start with. Even if we
perform $O(10^6)$ sweeps using standard Metropolis update with 3
reflection steps\footnote{The metropolis algorithm with reflection
  steps (MRS) is considered as a very effective local update algorithm
  for U(1) \cite{BUN94}.}, SCSD prevents a sufficiently accurate
determination of the phase weight.  Therefore, we install a recursive
procedure: from a previous guess $\hat{N}_{\mbox{\tiny
    i}}(S,\hat\beta^i_c)$ we go through MHMC and arrive at
$\hat{N}_{\mbox{\tiny i+1}}(S,\hat\beta^{i+1}_c)$. This computational
scheme is initialized by a standard canonical ``short run''.  We found
that one such learning cycle is sufficient.  \fig{WEIGHT} illustrates
the evolution of the multicanonical action density on the $16^4$
lattice.

\begin{figure}[!htb]
\centerline{\includegraphics[width=1.1\textwidth]{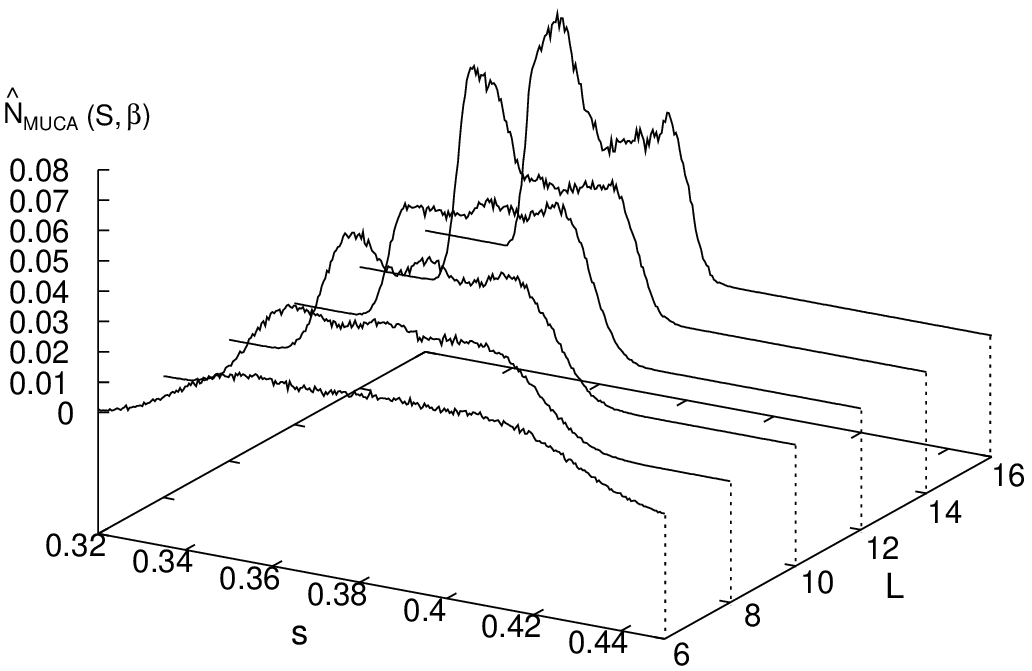}}
\caption{\label{FLAT}  
  $\hat N_{\mbox{\tiny MUCA}}(S,\beta)$ at run-parameter $\beta$ as
  given in  \tab{beta}.}
\end{figure}

On larger volumes the determination of a good guess can be
considerably boosted by a crank-up extrapolation that starts from
smaller systems \cite{BER97}.  In \fig{FLAT}, we display the quality
of ``flatness'' of $\hat N_{\mbox{\tiny MUCA}}(S,\beta)$ achieved in
our investigations for the various lattice sizes.

\subsection{Tunneling Behaviour}

\begin{figure}[!htb]
\centerline{\includegraphics[width=.7\textwidth]{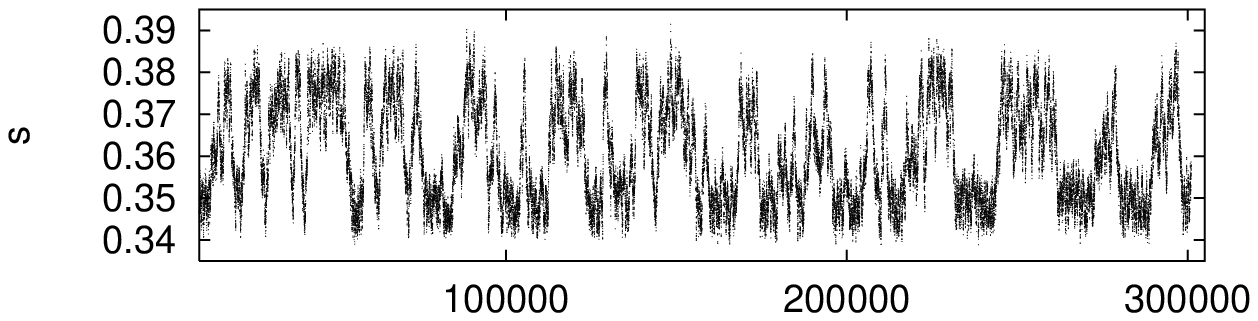}}
\centerline{\includegraphics[width=.7\textwidth]{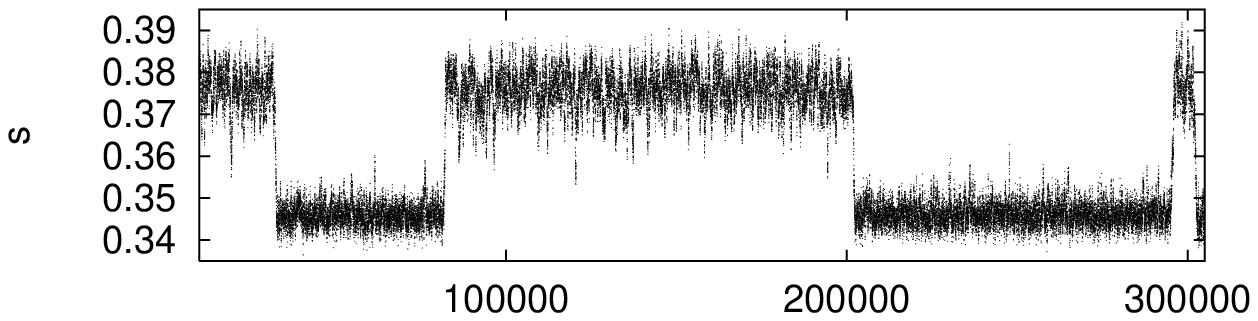}}
\caption{Time history of the $16^4$ system at $\beta=1.010753$
  for MHMC (top) and MRS (bottom).\label{series}}
\end{figure}

With our estimate for $\hat\nmuca^{16}(S)$ at $\beta=1.010753$,
depicted in \fig{WEIGHT}, we have generated the time history of the
action per site, $s=S/6V$, as shown in the upper part of \fig{series}.
For reference, we have included the time history from the MRS
algorithm on the same lattice. The figure demonstrates the success of
MHMC: the method provides us with a gain factor in tunneling rate of
about one order of magnitude on the $16^4$ lattice.

\begin{figure}[!htb]
  \centerline{\includegraphics[width=.8\textwidth]{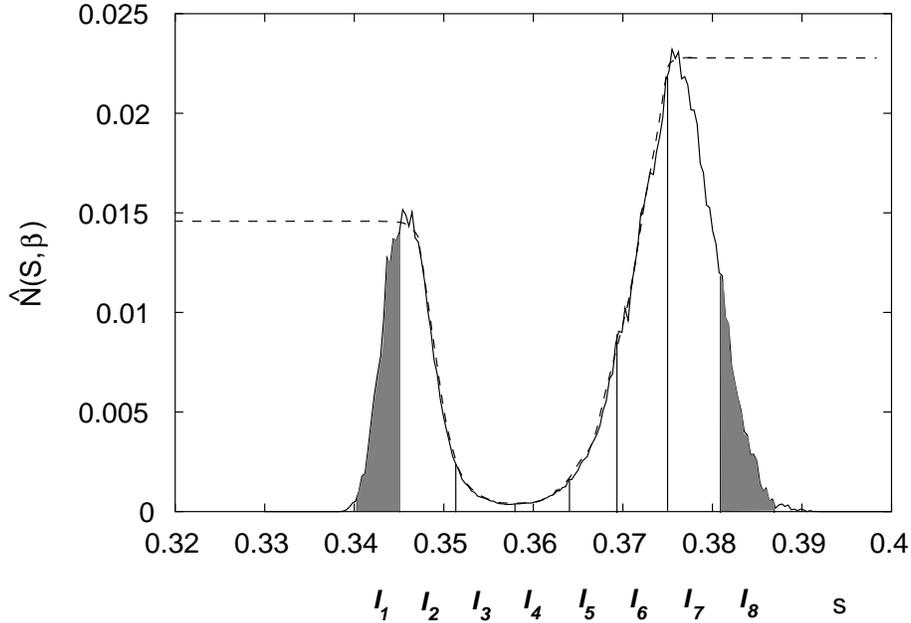}}
\caption{\label{plot16}  
  Action density of the $16^4$ system at ${\beta}=1.010753$ as a
  function of $s=\frac{S}{6V}$ from which $\hat\vmuca$ is derived (the
  dashed line shows $\exp(\hat\vmuca)$). The subdivision of the
  support of the action density into 8 intervals is introduced in
  order to define $\tauflip$.}
\end{figure}

In order to quantify this achievement, we introduce the average flip
time, $\tauflip$, a quantity that is readily measurable.  $\tauflip$
is defined as follows: we histogram the time series of $s$ using $N$
bins, as illustrated in \fig{plot16}.
A suitable number is $N=8$.  The total binning range is adjusted such
that $99.9$ \% of the events are covered symmetrically by the 8 bins.
A flip (flop) is given when the system travels from $I_8$ to $I_1$
(and vice versa).  $\tauflip$ is defined as the inverse number of the
sum of flips and flops multiplied by the total number of trajectories.  In
\tab{beta}, $\tauflip$ is given for the various lattices. The error in
$\tauflip$ has been computed by a jackknife error analysis.
\begin{table}[!htb]
\begin{center}
\begin{tabular}{|r|l|l|l|} 
\hline
L & $\beta$ & $\tauflip^{\mbox{\tiny MRS}}$ &
$\tauflip^{\mbox{\tiny MHMC}}$  \\ \hline\hline
6 & 1.001600 & 508(12) & 650(20)      \\ \hline
8 & 1.007370 & 1023(60) & 1173(50)    \\ \hline
10& 1.009300 & 2474(117)& 2006(84)    \\ \hline
12& 1.010143 & 5470(770) & 3260(440)   \\ \hline
14& 1.010668 & 16400(3300) & 5090(630) \\ \hline
16& 1.010753 & 44800(9700) & 6350(860) \\ \hline
\end{tabular}
\caption{\label{beta}
  $\tauflip$ for MRS and MHMC measured at the simulated $\beta$'s.}
\end{center}
\end{table}

\subsection{Scaling Behaviour}
 
\begin{figure}[!htb]
\centerline{\includegraphics[width=.7\textwidth]{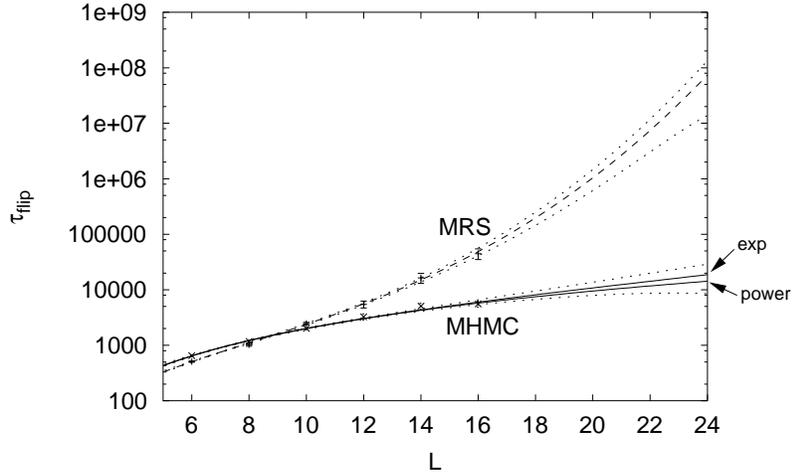}}
\caption{
  Tunneling times for MRS (exponential fit) and MHMC (lower convex
  curve is power law, upper convex curve is exponential fit). The
  errors of the two exponential fits are depicted as dotted lines.
  The error of the power law fit, \eq{powerfit}, is not visible on
  this scale.
  \label{algospeed}}
\end{figure}

With the results for $\tauflip$ on lattices up to $16^4$ we are in the
position to estimate the scaling behaviour of MHMC in comparison to
standard MRS updates.  \fig{algospeed} shows $\tauflip$ both for MHMC
and MRS as a function of the lattice size $L$ at their respective
$\beta$-values, as listed in \tab{beta}.

According to the expected exponential behaviour of
$\tauflip^{\mbox{\tiny MRS}}$ which, in the asymptotic regime $L \to
\infty$, should be given by \eq{EQ:SCSD}, we perform a $\chi^2$-fit
with the ansatz:
\begin{equation}
  \tauflip^{\mbox{\tiny MRS}}=a \ L^b \ e^{c L^3}.
\label{expfit}
\end{equation}
It yields the following parameter values:
\begin{eqnarray}
  a &=& 11.9(3.7)\nn\\
  b &=& 2.01(18) \nn\\
  c &=& 6.7(8)\ 10^{-4}, 
\end{eqnarray}
with $\chi_{\mbox{\tiny per
    d.o.f.}}^2=0.897$. As a result, we find a clear exponential SCSD
behaviour for the MRS algorihm\footnote{One is tempted to extract the
interfacial surface tension $\sigma$ from the fit to the MRS data. We find
\begin{equation*}%
\sigma=3.35(39) \ 10^{-4}.
\end{equation*}}.

On the other hand, for the tunneling times of the MHMC, we expect a
monomial dependence in $L$:
\begin{equation}
\tauflip^{\mbox{\tiny MHMC}}=p \ L^q,
\label{powerfit}
\end{equation}
We obtain for the fit parameters:
\begin{eqnarray}
p &=& 11.6(1.6)\\
q &=& 2.238(68). 
\end{eqnarray}
The power law ansatz is well confirmed by the fit quality with
$\chi_{\mbox{\tiny per d.o.f.}}^2=0.795$.  

We also took the pessimistic ansatz and tried to detect a potentially
exponential increase of $\tauflip^{\mbox{\tiny MHMC}}$.  The
exponential fit gives $\chi_{\mbox{\tiny per d.o.f.}}^2=0.975$.  As
can be seen in \fig{algospeed}, the exponential contribution remains
suppressed in the extrapolation.  A potentially dominating exponential
behaviour for MHMC can only be detected in future MHMC simulations on
larger lattices.  In other words, parallel MHMC {\em is} capable to
overcome SCSD in compact QED in practical simulations, at least up to
lattices sizes $\approx 24^4$.

\section{Cost Estimates for a  FSS Study.}

Finally, we try to assess the compute effort required to perform a FSS
study on a series of lattices ranging up to $24^4$.

 $\tauflip$ being readily accessible, we relate this quantity to the
{\em effective} integrated autocorrelation time, $\taueff$, defined in
Ref.~\cite{JAN94} by
\begin{equation}
\sigmamuca^2=\sigmacan^2\frac{2\taueff}{\Nts}.
\end{equation}
$\sigmamuca^2$ is the squared error of the observable $\cal O$
computed from the multicanonical ensemble, see \eq{AVE}.
$\sigmacan^2$ denotes the canonical variance of $\cal O$ (computed
from the reweighted canonical ensemble) and $\Nts$ is the length of
the multicanonical time series.  We can determine $\sigmamuca^2$ in a
numerically quite stable way from jackknife blocking.  From the time
series of $S$ on the $14^4$ and a $16^4$ MUCA system, with about
550000 and 300000 entries, respectively, we have determined $\taueff$
to be
\begin{eqnarray}
\taueff^{14}=35(10)\ 10^{-3} \times\tauflip^{14},\nn\\
\taueff^{16}=41(12)\ 10^{-3} \times\tauflip^{16}.
\end{eqnarray}
As a result, we found $\tauflip\approx
0.038(8)\times\taueff$\footnote{Note that $\tauflip$ strongly depends
  on the difference between $I_8$ and $I_1$ in \fig{plot16}. It
  remains to be confirmed that the ratio between $\taueff$ and
  $\tauflip$ does not vary too much going to larger lattices.}. From hereon
we can estimate the number of de-correlated subsamples (independent
measurements) out of a time series of length $\Nts$ to be roughly
\begin{equation}
  \Nindep=\frac{\Nts}{2\taueff}=\frac{\Nts}{79(16)\ 10^{-3}\tauflip}.
\end{equation}
Assuming the inverse square-root of $\Nindep$ to be an upper bound to
the relative error $r$ of an observable $\cal O$, we arrive at
\begin{equation}
  K=\frac{79(16)\ 10^{-3}}{r^2},
\end{equation}
with $K$ being the required number of flips to achieve a relative
error $<r$.

We thus conclude that $O(100)$ flip-flops might allow to determine
quantities like the specific heat and the Binder-Landau cumulant with a relative error of 3 \%.

Obviously, the costs of MHMC and MRS simulation increase with the
volume of the lattice, $V=L^4$.  Additionally, for MHMC, we want to
keep the average acceptance probability of the leapfrog scheme
constant.  To this end, we have to lower the step size according to
$\Delta t \sim V^{-1/4}$.  In a detailed tuning investigation we have
confirmed that the scaling rule of constant acceptance probability
\cite{CREUTZ88} leads to optimal performance. From a spectral analysis
of the molecular dynamics we can find an optimized trajectory length,
$N_{md}$, (in the sense that the average acceptance probability is
maximized) fulfilling $\Delta t \ N_{md}= \mbox{const}(\beta)$, with
only a slight dependence on $\beta$ near the phase transition.  We
choose step-size $\Delta t$ according to
\begin{equation} 
  \langle P_{\mbox{\tiny acc}} \rangle =
  \mbox{erfc}\Big(c(\beta)V^{\frac{1}{2}}\Delta t^2\Big)=\mbox{const.},
\end{equation}
adjusted such that the product $\tau_{\mbox{\tiny int}} N_{md}$ is
minimized finally. In our case, the optimal acceptance probability is
65 \% \cite{TUNE98}.

\begin{figure}[htb]
\centerline{\includegraphics[width=.7\textwidth]{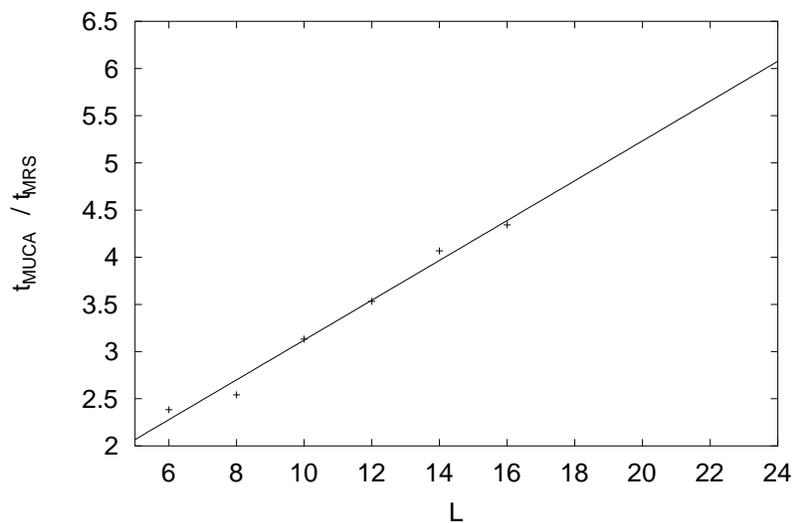}}
\caption{
  Ratio of CPU-times per sweep $\frac{t_{\mbox{\tiny
        MUCA}}}{t_{\mbox{\tiny MRS}}}$ and linear fit. Errors are not
  visible on this scale.  \label{cputime}}
\end{figure}

\fig{cputime} confirms the scaling of MHMC: the ratio of measured
CPU-times for a sweep, $\frac{t_{\mbox{\tiny MUCA}}}{t_{\mbox{\tiny
      MRS}}}$ increases quite linear with $L=V^{1/4}$. Minor
deviations from this behaviour stem from the use of suboptimal run
parameters $\Delta t$ and $N_{\mbox{\tiny md}}$.

The product $\tauflip\times t_{\mbox{\tiny MUCA}}$ reflects the
efficiency of MHMC.  \tab{gain} lists the effective gain factors
achieved taking MHMC instead of MRS.  The two columns refer to the
power law and exponential extrapolations for MHMC.
\begin{table}[!htb]
\begin{center}
\begin{tabular}{|r|l|l|} \hline
L &   power&  exponential \\ 
\hline\hline
16 & 1.9(3) & 1.9(3) \\ \hline
18 & 5.5(1.5) & 5.1(1.6) \\ \hline
20 & 21.0(8.7)& 18.4(9.0) \\ \hline
22 & 112(67) & 92(65) \\ \hline
24 & 856(700) & 652(638) \\ \hline
\end{tabular}
\caption{Gain factor for MHMC over MRS as function of the
  linear lattice extension.  The prediction for $\tauflip^{\mbox{\tiny
      MHMC}}$ is based on a power law ansatz (row 1) and an exponential
  ansatz (row 2). \label{gain}}
\end{center}
\end{table}

\begin{figure}[!htb]
\centerline{\includegraphics[width=.7\textwidth]{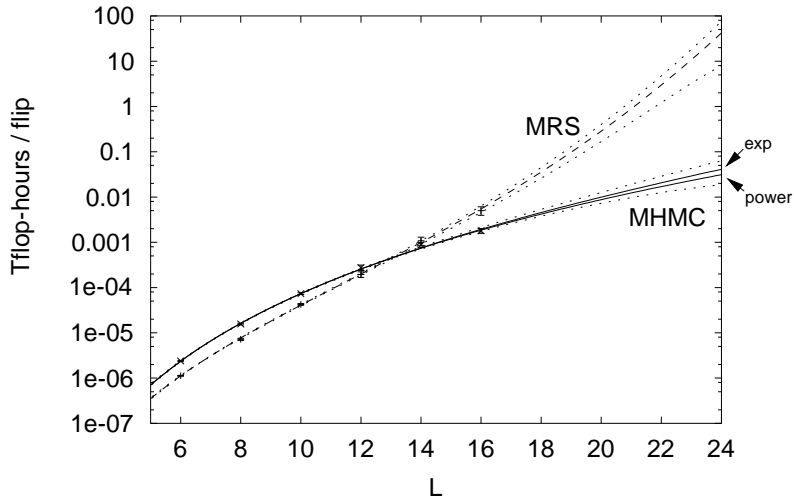}}
\caption{
  Sustained CPU time in Tflop-hours to generate one flip.
\label{cpuspeed} }
\end{figure}

Let us finally translate these factors into real costs: in \fig{cpuspeed},
we extrapolate the sustained CPU time in Tflop-hours required to
generate one flip. 
We conclude, that the integrated CPU time to generate the required 100
flips with MHMC on a $24^4$ lattice amounts to about 3 Tflop-hours
sustained CPU time.  

\section{Summary and Outlook}

We have demonstrated that the fully parallel MHMC algorithm is a very
effective tool which is able to overcome SCSD as present in the
pronounced metastabilities of 4-dimensional $U(1)$ gauge theory.  A
FSS study up to a lattice size of $L=24$ with about 100 flip events
for each lattice is feasible within half a year runtime, given a
sustained performance of about 2 Gflops, due to the improvements
achieved by MHMC. These performance figures should be obtainable on a
32 node partition of a Cray T3E-600.

Less well known is the influence of the delicate part of MHMC, \ie\ 
the determination of a suitable estimate for $\vmuca$, which is
carried out in an iterative manner. So far, we have encouraging
experiences on the $16^4$ lattice. The success of the crank-up
procedures described in Ref.~\cite{BER97} gives us hope that the
$\pmuca$ determination will carry through with only marginal
deterioration of the improvement factors estimated here.

The investigations presented form part of an ongoing study that aims at
a conclusive FSS analysis of compact QED on the Wilson line
\cite{ARN98}.

\section*{Acknowledgements}

The present investigation is based on the diploma thesis of G.\ A.\ 
carried out on the Connection machine CM5 at Wuppertal university.  We
thank P. Ueberholz and P. Fiebach for their friendly support.  We are
indebted to Thomas Riechmann and Dr.\ Claus-Uwe Linster of the
Institut f\"ur Mathematische Maschinen at the computer center of
Erlangen university, Germany, for providing us a substantial amount of
computer time on their 32 node connection machine CM5. Without this
support, the present investigation would not have been possible.


\begin{thebibliography}{99}
\frenchspacing
%
\bibitem{GUT78} A. H. Guth: \prd{21}{1980}{2291}.
%
\bibitem{MET53} N. Metropolis, A.W. Rosenbluth, M.N. Rosenbluth,
A.H. Teller and E. Teller: \jcep{21}{1953}{1087} 
%
\bibitem{REB79} M. Creutz, L. Jacobs, and C. Rebbi:
  \prd{20}{1979}{1915}.
%
\bibitem{ADL81} S. Adler: \prd{23}{1981}{2901}.
%
\bibitem{BUN94} B. Bunk: proposal for U(1) update, unpublished,
  private communication.
%
\bibitem{FIS70} M. E. Fisher and M. N. Barber: \prl{28}{1971}{1516}.
%
\bibitem{BER91} B. A. Berg and T. Neuhaus: \plb{267}{1991}{249}.
%
\bibitem{DUA87} S. Duane et al.: \plb{195}{1987}{216}.
%
\bibitem{LAU80} B. Lautrup and M. Nauenberg: \plb{95}{1980}{63}.
%
\bibitem{BHA81} G. Bhanot:  \prd{24}{1981}{461}.
%
\bibitem{MUE82} K.-H.\ M\"utter and K.\ Schilling,
  \npb{200}{1982}{362}.
%
\bibitem{GUP86} R. Gupta, A. Novotny, and R. Cordery:
  \plb{172}{1986}{86}.
%
\bibitem{JER83} J. Jers\'ak, T. Neuhaus, and P. M. Zerwas:
  \plb{133}{1983}{103}.
%
\bibitem{AZC91} V. Azcoiti, G. di Carlo, and A. Grillo:
  \plb{268}{1991}{101}.
%
\bibitem{BHA92} G.~Bhanot, Th.~Lippert, K.~Schilling, and
  P.~Ueberholz: \npb{378}{1992}{633}.
%
\bibitem{LAN87} C. B. Lang: \npbfs{280}{18}{1987}{255}.
%
\bibitem{HAS88} K. Decker, A. Hasenfratz, and P. Hasenfratz:
  \npbfs{295}{21}{1988}{21}.
%
\bibitem{KLA97} B. Klaus and C. Roiesnel: hep-lat 9801036
%
\bibitem{BHA81-2} G. Bhanot \npb{205}{1992}{168}
%
\bibitem{CAM97} I. Campos, A. Cruz, and A. Taranc\'on:
  \plb{B424}{1998}{328}.
%
\bibitem{CAM98} I. Campos, A. Cruz, and A. Taranc\'on:
  \npb{528}{1998}{325}.
%
\bibitem{EVE85} H. G. Evertz, J. Jers\'ak, T. Neuhaus, and P. M.
  Zerwas: \npb{251}{1985}{279}.
%
\bibitem{JER97} J. Cox, W. Franzki, J. Jers\'ak, C. B. Lang, T.
  Neuhaus, P. W. Stephenson \npb{499}{1997}{371}.
%
\bibitem{JER97-2} J. Cox, W. Franzki, J. Jers\'ak, C. B. Lang, T.
  Neuhaus, A. Seyfried, and P. W. Stephenson \npps{63}{1998}{691}.
%
\bibitem{JER85} V. Gr\"osch et al.: \plb{162}{1985}{171}.
%
\bibitem{LAN94} C. B.Lang and T. Neuhaus: \npps{34}{1994}{543}.
%
\bibitem{LAN94-2} C. B.Lang and T. Neuhaus: \npb{431}{1994}{119}.
%
\bibitem{JER95} J. Jers\'ak, C. B.Lang and T. Neuhaus:
  \npps{42}{1995}{672}.
%
\bibitem{BAI94} M. Baig and H. Fort: \plb{332}{1994}{428}.
%
\bibitem{LIP94} A. Bode, Th. Lippert, and K. Schilling:
  \npps{34}{1994}{1205}.
%
\bibitem{LIP95} Th. Lippert, A. Bode, V. Bornyakov, and K. Schilling:
  \npps{42}{1995}{684}.
%
\bibitem{LIP95-2} Th. Lippert, V. Bornyakov, A. Bode, and K.
  Schilling: {\it Monopoles in Compact U(1) - Anatomy of the Phase
    Transition}, in H. Toki, Y. Mizuno, H. Suganuma, T. Suzuki, O.
  Miyamura (edts.): International RCNP Workshop on Color Confinement
  and Hadrons (CONFINEMENT 95), Osaka, Japan, 22-24 Mar, 1995,
  Proceedings CONFINEMENT 95, World Scientific, 1995, pp. 247-254.
%
\bibitem{MAR92} E. Marinari and G. Parisi: Europhys. Lett. 19 (1992) 451.
%
\bibitem{KER94} W. Kerler, C. Rebbi, and A. Weber:
  \prd{50}{1994}{6984}.
%
\bibitem{KER94-2} W. Kerler, C. Rebbi, and A. Weber:
  \npb{450}{1995}{452}.
%
\bibitem{ADL91} S. L. Adler, G. Bhanot, Th. Lippert, K.  Schilling,
  and P. Ueberholz: \npb{368}{1992}{745}.
%
\bibitem{TORRIE76} G. M. Torrie and J. P. Valleau: \jcp{23}{1977}{187}.
%
\bibitem{TORRIE762} G. M. Torrie and J. P. Valleau:
  \jcep{66}{1977}{1402}.
%
\bibitem{VALLEAU90} J. P. Valleau: \jcp{91}{1996}{193}.
%
\bibitem{DING92} K. Ding and J. P. Valleau: \jcep{98}{1993}{3306}.
%
\bibitem{BER91-2} B. A. Berg and T. Neuhaus: \prl{68}{1992}{9}.
%
\bibitem{BER95} B. A. Berg: \jsp{82}{1996}{323}.
%
\bibitem{BER97} B. A. Berg: Proceedings of the International
  Conference on Multiscale Phenomena and Their Simulations (Bielefeld,
  October 1996), edited by F. Karsch, B. Monien and H. Satz (World
  Scientific, 1997).
%
\bibitem{BER97-2} B. A. Berg: \npps{63}{1998}{982}.
%
\bibitem{GRO92} B. Grossmann et al.: \plb{293}{1992}{175}.
%
\bibitem{RUM92} K. Rummukainen: \npb{390}{1993}{621}.
%
\bibitem{JAN93} W. Janke and T. Sauer: \npps{34}{1994}{771}.
%
\bibitem{JAN93-2} W. Janke and T. Sauer: \pre{49}{1994}{3475}.
%
\bibitem{HAN94} U. H. E. Hansmann and Y. Okamoto: hep-lat/9411005.
%
\bibitem{NEU96} T. Neuhaus: heplat/9608043, preprint Wuppertal WUB
  96-30.
%
\bibitem{ARN98} G. Arnold, Th. Lippert, and K. Schilling: to appear.
%
\bibitem{GOT87} S. A. Gottlieb, W. Liu, D. Toussaint, R. L. Renken,
  and R. L. Sugar: \prd{35}{1987}{2531}.
%
\bibitem{LIP97} Th. Lippert in: H. Meyer-Ortmanns and A. Kl\"umper
  (edts.): Proceedings of {\it Workshop of the Graduiertenkolleg at
    the university of Wuppertal on `Field Theoretical Tools for
    Polymer and Particle Physics'} (Springer, Berlin, 1998), p. 122.
%
\bibitem{SEYFRIED} A. Seyfried, doctorate thesis, Wuppertal
  University, 1998, unpublished and private communication.
%
\bibitem{SF} A. M. Ferrenberg, R. H. Swendsen: \prl{61}{1988}{2635}
  and \prl{63}{1989}{1658}.
%
\bibitem{JAN94} W. Janke and T Sauer: \jsp{78}{1995}{759}.
%
\bibitem{CREUTZ88} M. J. Creutz, {\it 'Quantum Fields on the
    Computer'}, Vol. 11, Advanced Series on Directions in High Energy
  Physics (World Scientific, Singapore, 1992).
%
\bibitem{TUNE98} G. Arnold, Th. Lippert, and K. Schilling: to appear.
%
\nonfrenchspacing 
\end{thebibliography}
\end{document}